\documentclass[]{aa}  
\bibpunct{(}{)}{;}{a}{}{,} 
\usepackage{graphicx}
\usepackage{txfonts}
\begin{document}

   \title{The puzzle of the CNO isotope ratios in AGB carbon stars}

   \author{C. Abia\inst{1}
          \and
          R. P. Hedrosa\inst{2}
          \and
          I. Dom\'\i nguez\inst{1}
          \and 
          O. Straniero\inst{3}
          }

   \institute{Dpto. F\'\i sica Te\'orica y del Cosmos, Universidad de Granada, E-18071 Granada, Spain
              \email{cabia@ugr.es}
         \and 
            Observatorio Astron\'omico Calar Alto, 04550 Gergal (Almer\'\i a), Spain
         \and
             INAF, Osservatorio Astronomico di Collurania, 64100 Teramo, Italy             
             }

   \date{Received; accepted }

 
  \abstract
   {The abundance ratios of the main isotopes of carbon, nitrogen and oxygen are modified by the CNO-cycle in the stellar
interiors. When the different dredge-up events mix the burning material with the envelope, valuable information 
on the nucleosynthesis
and mixing processes can be extracted by measuring these isotope ratios.}
   {Previous determinations of the oxygen isotopic ratios in asymptotic giant branch (AGB) carbon stars were at odds with 
the existing theoretical predictions. We aim to redetermine the oxygen ratios in these stars using new spectral analysis 
tools and further develop discussions on the carbon
and nitrogen isotopic ratios in order to elucidate this problem.}
   {Oxygen isotopic ratios were derived from spectra in the K-band in a sample of galactic AGB carbon stars of different spectral types 
and near solar metallicity. Synthetic spectra calculated in local thermodynamic equillibrium (LTE) with spherical carbon-rich 
atmosphere models and updated molecular line lists 
were used. The CNO isotope ratios derived in a homogeneous way, were compared with theoretical predictions for 
low-mass ($1.5-3$ M$_\sun$) AGB stars computed with the FUNS code assuming extra mixing both during the RGB and AGB phases.}
   {For most of the stars the $^{16}$O$/^{17}$O$/^{18}$O ratios derived are in good agreement with theoretical predictions 
confirming that, for AGB stars, are established using the values reached after the first dredge-up (FDU) according
to the initial stellar mass. This fact, as far as the oxygen isotopic ratios are concerned, leaves little space for the operation of 
any extra mixing mechanism 
during the AGB phase. Nevertheless, for a few stars
with large $^{16}$O$/^{17}$O$/^{18}$O, the operation of such a mechanism might be required, although their observed $^{12}$C$/^{13}$C 
and $^{14}$N$/^{15}$N ratios would be difficult to reconcile within this scenario. Furthermore, J-type stars 
tend to have lower $^{16}$O$/^{17}$O ratios than the normal carbon stars, as already indicated in previous studies. Excluding these peculiar 
stars, AGB carbon stars occupy the same region as pre-solar type I oxide grains in a $^{17}$O$/^{16}$O vs. $^{18}$O$/^{16}$O diagram, showing
little spread. This reinforces the idea that these grains were probably formed in low-mass stars during the previous O-rich phases.} 
   {}


   \maketitle
%

\section{Introduction}
Asymptotic giant branch (AGB) stars are one of the major chemical polluters of
the interstellar medium (ISM) in the galaxies. It is believed that more than 50$\%$ of
the material returned to the ISM by dying stars comes from AGB stars \citep[e.g.,][]{gos14}, which constitute
the late phase in the evolution of low-and intermediate-mass ($1\lesssim\rm{M/M}_\sun\lesssim 8$) stars.
AGB stars also play  a key role in the formation of the ISM dust as many grain species form in their
cool circumstellar envelope \citep[e.g.][]{geh89}. Many of these interstellar grains present in the
protosolar nebula and preserved in meteorites, can now be analyzed in the laboratory
\citep[see e.g.,][]{dav11} with unprecedented accuracy. The composition of the AGB ejecta depends on the initial composition and on the nuclear
and mixing processes occurring during the stellar lifetime. Of special interest in this sense are 
the isotopic ratios of carbon ($^{12,13}$C), nitrogen ($^{14,15}$N), and oxygen ($^{16,17,18}$O).

Stars in the red giant branch (RGB) undergo the so-called 
first dredge-up (FDU), a convective mixing process that carries nuclei from internal layers, 
previously affected by partial CNO cycling to the surface. It is well established  that this
leads to a  decrease in the  atmospheric $^{12}$C/$^{13}$C ratio with respect
to the main sequence (MS) value  ($\sim 89$ in the  solar  case), down to values
in the range $20-30$ \citep[depending  on the initial  mass and
metallicity of  the star, see e.g.] []{wei00}.  In addition, the carbon abundance drops in the envelope, while that
of $^{14}$N increases and $^{15}$N is depleted, so the $^{14}$N/$^{15}$N ratio increases respect to the initial
value \citep[459 in the solar case,][]{mar11}. The $^{16}$O abundance remains unaltered, while that
of $^{17}$O increases by $\sim 50\%$ and $^{18}$O is mildly reduced. The  oxygen isotopic ratios
predicted by the models after the FDU are modified with respect to the solar values 
\citep[$(^{16}\rm{O}/^{17}\rm{O})_\sun=2700$  and $(^{16}\rm{O}/^{18}\rm{O})_\sun=498$,][]{lod09}; 
their values depend on the initial stellar
mass and metallicity and have changed strongly in recent years as a consequence of
changes in basic CNO cycle reaction rates. The present situation for the CNO 
ratios after the FDU as a function of the stellar mass, updated with the last version of the FUNS evolutionary code 
\citep{stra06} and with the last recommendations available for the relevant reaction rates
\citep{ade11,boe16}, is summarized in Table \ref{t1}. 

A second dredge-up (SDU) takes place in  stars
with M$> 4$ M$_\sun$ during the early-AGB phase bringing to the surface fresh $^{4}$He and
$^{14}$N mainly, although the CNO isotopic ratios in the envelope are not significantly modified.
During the main AGB phase a third dredge-up (TDU) occurs when the convective envelope penetrates inward into
the H-exhausted core after a thermal pulse (TP). Products of the He burning are mixed into the envelope
as a consequence of the recurrent TDU episodes. $^{12}$C is the main product of the He burning so that 
the $^{12}$C$/^{13}$C ratio is expected to increase significantly in the envelope during the TP-AGB phase from the values 
previously set by the FDU. Eventually, the amount of carbon in the envelope might exceed that of oxygen (C/O$>1$, by number), and
then the star becomes an AGB carbon star. Standard AGB models indicate that carbon stars are
formed between masses 1.5-3 M$_\sun$\footnote{These limits depend on the pre-AGB phase mass loss and on the initial
chemical composition.}. In models with initial mass M$<1.5$ M$_\odot$, the envelope mass at the beginning of
the AGB phase is quite small, so that only a few thermal pulses will be experienced before the end of the
AGB. As a consequence the amount of carbon eventually dredged-up is not enough to attain the carbon star condition.
Models also show that the $^{14}$N/$^{15}$N and $^{16}$O/$^{17}$O/$^{18}$O 
ratios are barely modified by TDUs \citep[see][for a review]{ir83,kar14,cri15}, so the expected N and O isotopic ratios at the surface
of an AGB star would be basically those after the FDU. Two occurrences prevent AGB stars 
with mass larger than 3-4 M$_\sun$ to become carbon stars \citep[see, e.g.,][]{str14}. First, the larger 
the core mass the weaker the TDU. Second, the temperature at the base of the convective envelope may be high enough 
to convert C in to N. It is the well-known Hot Bottom Burning \citep[see][]{ir83}. In this case, important 
modification of the C, N and O isotopic ratios are expected \citep[see e.g.][and references therein]{ven15} as well as of other
light elements (Li, Na, Mg, Al). 
\begin{table*}
\caption{\label{t1} Predicted CNO ratios (by number) after the FDU at solar metallicity (Z$=0.014$, Y$=0.27$).}
\centering
\begin{tabular}{lccccc}
\hline\hline
Mass (M$_\sun$) &$^{12}$C$/^{13}$C &$^{14}$N$/^{15}$N& $^{16}$O$/^{17}$O& $^{16}$O$/^{18}$O & $^{17}$O$/^{18}$O\\
\hline
1.2 & 29 & 576 & 2192 & 583 & 0.26 \\
1.5 & 26 & 782 & 1095 & 637 & 0.58 \\
1.8 & 25 & 974 & 453 & 681 & 1.50 \\
2.0 & 24 & 1071 & 294  & 700 & 2.38 \\
2.5 & 24 & 1316 & 273  & 709 & 2.59 \\
3.0 & 24 & 1458 & 323  & 708 & 2.19 \\
4.0 & 23 & 1548 & 406  & 703 & 1.73 \\
\hline
\end{tabular}
\end{table*}
\subsection {The need of extra mixing}

Nevertheless, this standard picture is challenged by a large amount of abundance
determinations \citep[see for example][]{bro89,gra00,char04,gru04} in
field and globular cluster, low-mass ($<3$ M$_\sun$) RGB stars. These stars show very low
$^{12}$C/$^{13}$C ratios, sometimes almost  reaching the
equilibrium value of the CN cycle ($\sim 3.5$). Anomalies in the
C and O isotopes are also found in pre-solar C-rich and O-rich
grains of stellar origin \citep[e.g.,][]{ama01,nit08}.
It became common to attribute these chemical anomalies to the occurrence of episodes of
matter circulation and/or diffusive processes \citep{was95,nol03,pal11x,denis98,egg2006}. These phenomena, 
known under the generic name of 
extra mixing, would link the envelope to regions where proton capture takes place,
thus accounting for the observation that the envelope material has undergone 
extensive nuclear processing. The specific mechanism triggering this extra mixing is not known, although
rotational (meridional) mixing, magnetic buoyancy, gravitational waves, and thermohaline circulation have been proposed
\citep{nor08,cha10,nuc14}. The existence of this non-standard mixing is widely accepted (whatever is its nature) 
after the so-called {\it bump} in the luminosity function 
during the RGB phase, but its operation on the AGB phase is highly debated \citep{leb08,kar10,bus10}. 
In fact, the only observational evidence is the low $^{12}$C$/^{13}$C ($<30$) ratios found \citep{olo93,abi02,mil09} 
in a significant fraction ($\sim 30\%$) of Galactic AGB carbon stars.
This poses a problem   since   standard   AGB   models   predict   a   minimum   of
$^{12}$C/$^{13}$C$\sim 30$ at C/O$>1$,  that is, at the C-rich AGB phase for
solar metallicity. This prediction is very robust as it only depends on the assumption that
$^{16}$O and $^{13}$C content in the envelope does not change due to
operation of  the TDU, which is  actually the case \citep[see figures in][]{abi111,hin16}. 
Even assuming that these stars arrive to the AGB phase with a low carbon ratio ($(^{12}$C/$^{13}$C$)_{\rm{RGB}}\sim 12$)  
as commonly observed in bright RGB stars in the Galactic disk, the minimum value predicted on the AGB at C/O$=1$ is 
$^{12}$C/$^{13}$C$\sim 30$. Note that this is independent of the efficiency of the TDU and
other model assumptions. 

A valuable piece of information on this discussion is provided by 
the $^{14}$N/$^{15}$N ratio. This is because the operation of the CNO cycle easily destroys $^{15}$N by
proton capture reactions while it simultaneously produces $^{14}$N. Thus, theoretically, the occurrence of extra mixing 
episodes during the RGB or/and AGB, would increase the $^{14}$N/$^{15}$N from the value typically attained after
the FDU ($\sim 1000$ assuming a solar initial ratio, see Table 1). Recently, \citet{hed13} 
measured this isotopic ratio for the first time in a sample of Galactic AGB carbon stars. Despite 
the large uncertainties, the nitrogen ratios found for normal (N-type) carbon stars lay above 
$^{14}$N/$^{15}$N$\gtrsim 1000$. As their detection limit was $^{14}$N/$^{15}$N$\lesssim 5000$, 
these authors cannot exclude the existence of carbon stars with higher nitrogen ratio  but
then an anti-correlation between nitrogen and carbon isotopic ratios would be expected, which is not observed
\citep[see figure 2 in][]{hed13}. This figure is in agreement with standard theoretical models 
for low-mass TP-AGB stars \citep[e.g.,][]{cri15} and, therefore, disfavors the operation of extra mixing 
during the AGB phase. 

Additional information can be extracted from the $^{16}$O/$^{17}$O/$^{18}$O ratios in red giants. 
Oxygen ratio determinations in field low-mass RGB stars show a general agreement with standard theoretical
predictions after the FDU \citep[see e.g.,][]{har88,smi89,smi90}, although the expected dependence on the stellar mass
(see Table 1) has not been fully confirmed, mainly because the large uncertainties in the observations and determination of the stellar  masses.
Attempts to do this in bright Galactic RGB stars with well-known stellar parameters or in giant stars belonging to
globular clusters with accurate ages and distances (thus, turn-off mass) show, nevertheless, a comfortable agreement with stellar models;
requiring, in some cases, the existence of extra mixing \citep{abi12} while not in others \citep{leb15}. On the other hand, previous 
determinations of the oxygen isotopic ratios in AGB stars date from the late 80's 
\citep{dom86,har87,smi90}. In particular, \citet{har87} (hereafter H87) derived oxygen ratios in carbon stars of N-, 
SC- and J-types. For ordinary N-type stars they found $550\leq^{16}$O$/^{17}$O$\leq 4100$ and 
$700\leq^{16}$O$/^{18}$O$\leq 2400$, and for J- and SC-type stars a tendency to show 
lower $^{16}$O$/^{17}$O ratios than the N-type was found. Also, they discovered
correlation between the $^{16}$O$/^{17}$O ratio and the neutron exposure\footnote{Theoretically, 
the neutron exposure $\tau=\int N_{n}v_{th}dt$, where $N_{n}$ is the neutron density and $v_{th}$ the 
thermal velocity, is often used to evaluate the s-process efficiency \citep[see][]{cla68}. Observationally,
the neutron exposure can be measured from the ratio [hs/ls], where hs represent the high-mass $s$-elements 
(Ba, La, Ce, Nd, and Sm) and ls the low-mass (Sr, Zr and Y). Here we adopt the definition 
[A/B]$=$log$(N_A/N_B)-$log$(N_A/N_B)_{\odot}$ where ($N_X$) is the abundance by number 
of the element $X$ in the scale $N_H\equiv 12$.}. These results 
were extremely difficult to explain in terms of the 
stellar evolution models at that time. Such large $^{16}$O/$^{17}$O and $^{16}$O/$^{18}$O 
ratios ($>1500$) are found however, in some pre-solar oxide grains (those of group I and II, see WUSTL Pre-Solar Database at 
http://presolar.wustl.edu/$\sim$pgd/), which are thought to form in oxygen-rich environments \citep{hus92} 
such as RGB and AGB stars with M$< 2$ M$_\sun$. In fact, \citet{pal13} showed that the oxide grains 
depleted in $^{18}$O (thus with large $^{16}$O/$^{18}$O ratios) can be explained assuming some kind of 
extra mixing both during the RGB and AGB phases. Very recently, \citet{hin16} have derived oxygen ratios
in O-rich AGB stars of Mira and SRa variability types. The $^{16}$O$/^{17}$O ratios derived are within a large range 
200 to 7000, probably indicating a wide range in the initial masses ($1-3$ M$_\sun$) of their sample stars (see Table 1) 
rather than evidence of extra mixing: indeed the observed $^{17}$O-depleted Miras do not show the corresponding $^{12}$C- and $^{18}$O-depletion
(see discussion below). They also found some stars with excess $^{18}$O ($^{16}$O$/^{18}$O$<500$). Since these stars are
younger than the Sun, these authors ascribed this fact to a 
galactic chemical evolution effect: $^{18}$O is a secondary element thus the  $^{16}$O$/^{18}$O ratio should decrease with time
(increasing metallicity). 

It is evident that the situation outlined above concerning the existence of non-standard mixing processes 
during the AGB is far from conclusive. This is in part due to the existing inhomogeneous observational data coming from different analyses
using different techniques. In this study we have attempted to elucidate this situation through a joint analysis and discussion
of the CNO isotopic ratios derived in AGB carbon stars in a homogeneous way. To do that we have redetermined
oxygen isotopic ratios in a sample of Galactic carbon stars, many of them already studied
by H87. The results are discussed together with the carbon and nitrogen isotopic ratios in the same stars derived previously using 
equal spectral analysis tools. In the following sections we discuss the analysis of the
infrared spectra used to derive the oxygen ratios and then discuss all CNO isotopic ratios in the framework of
the state-of-the-art of stellar and nucleosynthesis models for AGB stars with and without RGB and/or AGB extra mixing.

\section{Observations and analysis}
The spectra used here were drawn entirely
from the archives of the Kitt Peak National Observatory
(KPNO) 4 m Mayall Telescope Fourier Transform Spectrometer
(FTS) or kindly provided by K. Hinkle. This material
has been extensively used in the literature for different purposes.
Our analysis is based on 2.3 $\mu$m spectra, in which lines due
to $\Delta v=2$ transition in the CO molecule's ground electronic
state are present. Indeed, many lines of the species $^{12}$C$^{16,17,18}$O
can be identified allowing the determination of the oxygen isotope ratios. 
At $\sim2.3$ $\mu$m these spectra have a resolving power of $R\sim 50000$. Details
on spectral reduction, calibration, and cleaning from telluric lines
can be seen in \citet{lam86} and H87 \citep[or in][for a recent summary]{hin16}, and will not be repeated here.
 
%

%
\begin{table}
\caption{\label{t2}Spectroscopic parameters of the CO lines used.}
\centering
\begin{tabular}{lcc}
\hline\hline
$^{12}$C$^{17}$O&      &           \\
\hline
Wavelength ({\AA})  &  $\chi$(eV)         &  log $gf$ \\
\hline  
23293.708 & 0.24 & -5.459 \\
23309.863 & 0.21 & -5.492 \\
23327.600 & 0.18 & -5.526 \\
23336.800 & 0.17 & -5.545 \\
23357.178 & 0.15 & -5.582 \\
23390.326 & 0.11 & -5.644 \\
23402.170 & 0.10 & -5.666 \\
\hline   
$^{12}$C$^{18}$O&    &  \\
\hline
23486.605 & 0.57 & -5.234 \\
23486.711 & 0.67 & -5.190 \\ 
23492.840 & 0.48 & -5.281 \\
23502.300 & 0.84 & -5.117 \\
23639.334 & 0.13 & -5.618 \\
23650.494 & 0.12 & -5.639 \\
23785.335 & 0.72 & -4.814 \\
23795.024 & 0.66 & -4.852 \\ 
23799.058 & 0.64 & -4.865 \\
23799.281 & 0.03 & -5.935 \\
23976.376 & 0.35 & -5.224 \\
23989.236 & 0.34 & -5.248\\
\hline
\end{tabular}
\tablefoot{Wavelengths are in air.}
\end{table}
%
  \begin{figure*}
   \centering
            \includegraphics[angle=-90,width=19cm]{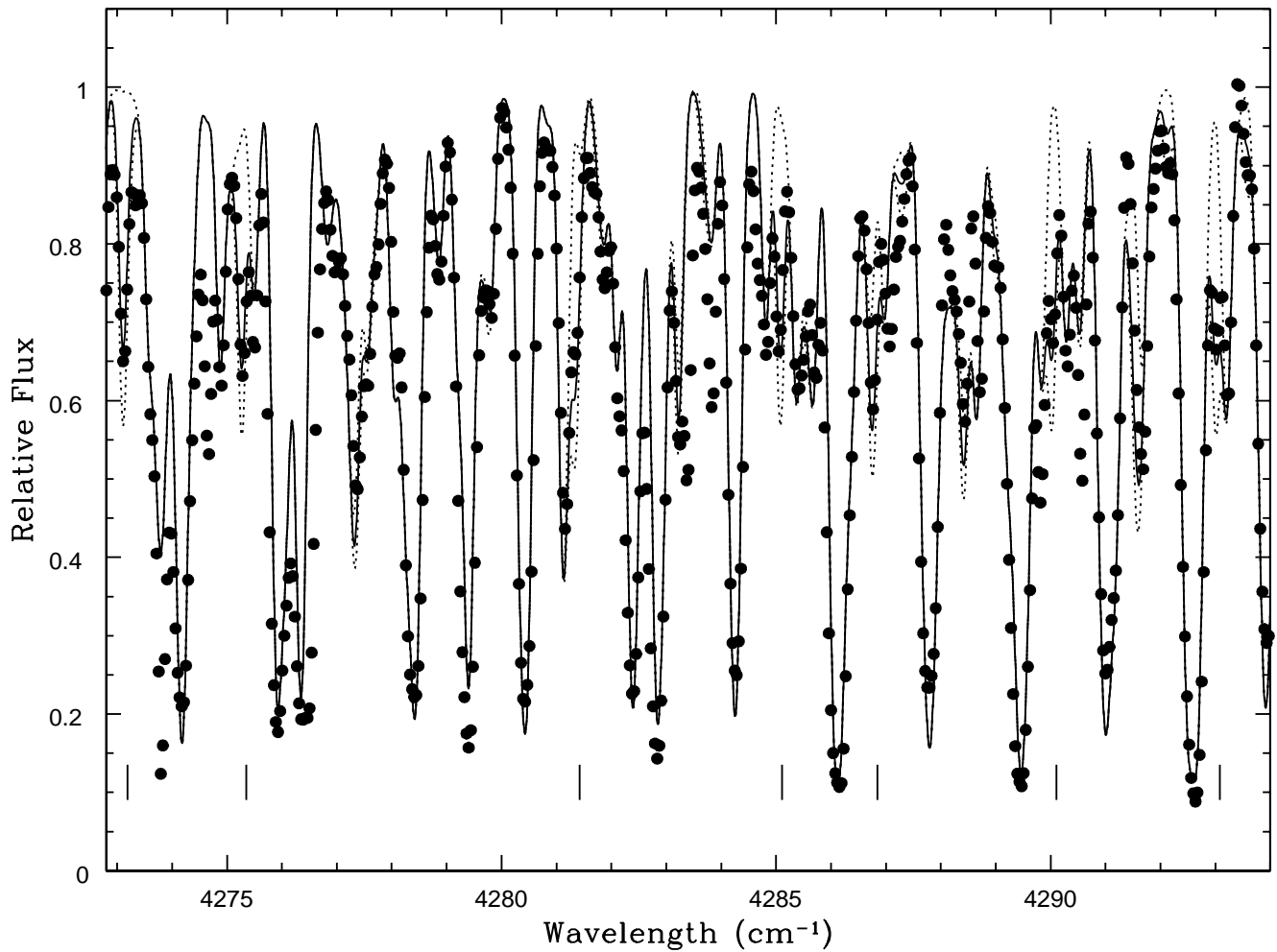}
      \caption{Comparison of observed (dots) and synthetic spectra for \object{Z Psc} (N-type). Synthetic spectrum
calculated with a $^{16}$O$/^{17}$O ratio 650 (continuous line). Synthetic spectra computed without $^{17}$O and
with $^{16}$O$/^{17}$O of 350, respectively (dotted lines). The position of some $^{12}$C$^{17}$O blends are marked with vertical lines.}
   \end{figure*}  

To perform a homogeneous analysis, we used the stellar parameters (T$_{eff}$, gravity, metallicities, microturbulence and C/O) previously derived
in our analysis of the fluorine abundances and the $^{14}$N$/^{15}$N ratios in the same stars 
\citep{abi10,hed13,abi15}. Of particular relevance is the determination of the CNO abundances. Carbon and 
oxygen abundances were derived from a few weak, unblended C$_2$ and CO lines in the 2.3 $\mu$m region. For most of the
stars, nitrogen abundances were derived from CN lines using very high-resolution spectra in the $\sim 8000$ {\AA} region
\citep[see][for details]{hed13}. Once the nitrogen abundance was determined, carbon and oxygen abundances were
again redetermined from the $2.3$ $\mu$m region in an iterative process until convergence was achieved. In the stars with
no spectrum in the 8000 ${\AA}$ region, we assumed the solar nitrogen abundance  according to \citet{aps09}. In any case, the
nitrogen abundance has a minor role on the derivation of the $^{16}$O/$^{17}$O/$^{18}$O ratios. AGB carbon
stars show typically solar nitrogen abundances \citep{lam86}. Updated atomic and molecular line lists were used
in both spectral regions \citep[see][for details]{hed13,abi10,abi15}.  A significant change with respect to
the line lists used in previous works is the updated CN line list. Furthermore, we have included 
the contribution from the HCN molecule (H$^{12}$CN/H$^{13}$CN) according to the computations by \citet{har03}. This molecule 
has some contribution by introducing a global extra absorption ($veil$) that might diminish the spectral continuum by up 
to $\sim 2-3\%$.  Careful selection of the $^{12}$C$^{17,18}$O lines in the
2.3 $\mu$m region was performed; in particular, we selected only weak lines not significantly affected
by blends, and sensitive to $^{17,18}$O abundance variations. The final list selected is shown in Table 2. Note that some of 
them are actually blends, which  may have an impact on the accuracy of the measured ratios. In a few stars, C$^{18}$O lines 
might be affected by telluric absorption not perfectly removed in the data reduction procedure. These lines were discarded 
when the $^{18}$O abundance was derived.  A C-rich spherical MARCS \citep{gus08} model atmosphere 
was chosen for each star according to its stellar parameters, and synthetic LTE spectra were calculated in
the $2.3$ $\mu$m and 8000 ${\AA}$ regions, by using the Turbospectrum v14 code \citep{ple12}. Theoretical spectra
were convolved with a Gaussian function with the corresponding full width to medium high to mimic the spectral line profile in each range 
which includes the macroturbulence parameter ($9-13$ kms$^{-1}$). We used $\chi^2$ minimization techniques to determine the 
oxygen ratios providing the best fit to each C$^{17,18}$O feature. The goal was to fit not only the selected lines, but 
also the
overall shape of the spectra. The final estimates of the CNO abundances  agree well with the results obtained
by \citet{abi10} and \citet{hed13} and may be taken as confirming them. The oxygen isotopic ratios derived from individual features were 
then combined to obtain 
an average. The number of spectral features used in each star to obtain the oxygen isotope ratios is given in Table 3.

%
   \begin{figure}
   \centering
   \includegraphics[width=9cm]{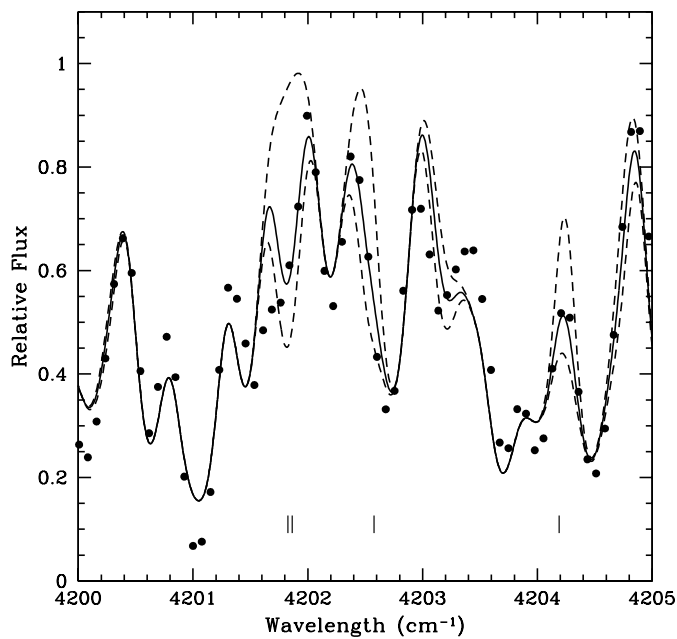}
      \caption{Comparison of observed (dots) and synthetic spectra for \object{FU Mon} (SC-type). The continuous line
represents the synthetic spectrum
calculated with a $^{16}$O$/^{18}$O ratio of 400. Dashed lines represent synthetic spectra computed without $^{18}$O and
with $^{16}$O$/^{18}$O of 200, respectively. The position of some $^{12}$C$^{18}$O blends are marked with vertical lines. Note
that some spectral features are not well reproduced by the synthetic spectra.}
         
   \end{figure}

%
   \begin{figure}
   \centering
   \includegraphics[width=9cm]{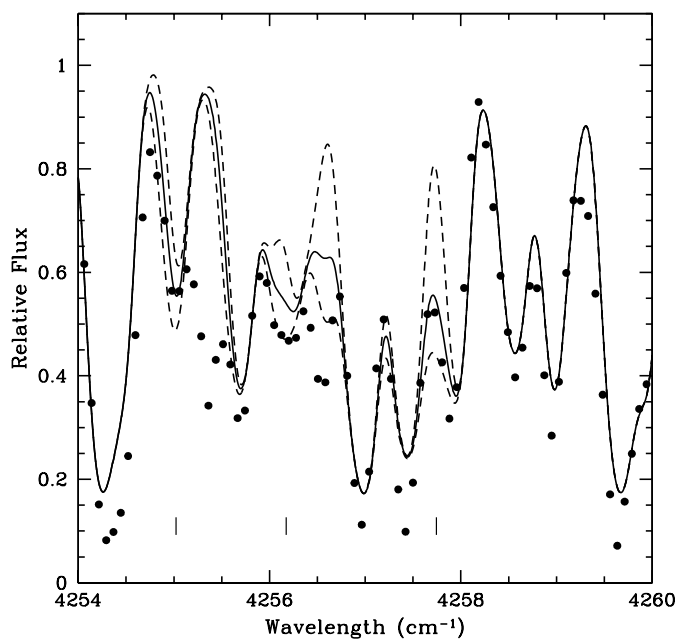} 
      \caption{As figure 2 for \object{FU Mon} in the vicinity of the $^{12}$C$^{18}$O 2-0 band-head.}
         
   \end{figure}

Figures 1 to 3 show examples of synthetic fits to observational spectra in the 2.3 $\mu$m region. 
Note that some spectral features are not well reproduced by the theoretical spectra in particular in the region 
where the C$^{18}$O features are present. Most of these lines are $^{12}$C$^{16}$O absorptions, which may indicate
that our line list is not yet complete. This certainly has a deleterious effect on the accuracy of 
measured $^{16}$O/$^{18}$O ratios. 

The derived oxygen ratios are affected by a number of uncertainty sources. In
particular the uncertainty in the C/O ratio because it has an important effect on the opacities in the model atmosphere. 
In fact, a typical uncertainty $\Delta$C/O$=\pm 0.02$ introduces an error in the oxygen ratios of $\sim 20\%$. The most important source 
of error is, however, the oxygen abundance. A change of $\pm 0.3$ dex in the [O/H] ratio may change the 
derived isotope ratios by up to $35\%$. Also, due to the almost saturation of the C$^{17,18}$O lines used, the 
assumed microturbulence ($\sim 2.3$ kms$^{-1}$) has an impact:
a change of $\Delta\xi=\pm0.3$ kms$^{-1}$ modifies the oxygen ratios by $\sim 15\%$. On the contrary, 
changes of $\pm 200$ K in the adopted effective temperature modifies the $^{16}$O$/^{17}$O$/^{18}$O ratios by only 
a few percent. Errors due to uncertainties in other model atmosphere parameters 
(gravity, metallicity, or nitrogen abundance) are even smaller. However, the estimate of the total
error is more complex than the simple quadratic addition of all these errors, because most of the stellar parameters 
are coupled to at least one of the others. For instance, a change
within the quoted uncertainty in the adopted T$_{eff}$ would change the oxygen ratios derived by a certain amount, but the resulting 
fit to the global spectrum might not be acceptable. Therefore, such a variation in the T$_{eff}$ value should be discarded. In other
words, the co-variance terms in the estimate of the total error have to be taken into account. In our case, and given the
inter-dependence between the stellar parameters, this is rather complex to estimate. To take this into account, 
a procedure designed  by \citet{cay04} was followed. Briefly, for
a typical star in the sample, a synthetic spectrum is computed by modifying a stellar parameter in an amount according to
its assumed uncertainty. Then, all the other stellar parameters are changed within their uncertainties until a good fit to
the global spectrum is found. The difference between the oxygen ratios derived using the first (best) adopted set of stellar
parameters and those derived with the new set of 'good' stellar parameters may give an estimate of the real uncertainty. 
The total error would then be that of the largest difference in the oxygen ratios obtained with the different combinations 
of the stellar parameters. We have performed this analysis for several stars in the sample. The error estimated in this way, 
added quadratically 
with that due to the uncertainty in the continuum position ($\sim 2\%)$ and the dispersion in the isotope ratios 
when derived from different lines (see Table 3), would give an estimate of the total error. This 
ranges from 250 to 350 and 360 to 550 for 
the $^{16}$O$/^{17}$O and $^{16}$O$/^{18}$O ratios, respectively (the larger the derived ratio the larger the total error). 
For J-type stars, the total uncertainty would be higher because synthetic fits to the observed spectra are, 
in general, of worse quality. In fact, for
some of these stars, we set only upper limits to the $^{18}$O abundance.

Now we compare  our results with previous estimates of the oxygen isotope ratios in the literature. Most of our stars are
in common with H87. We find systematically lower $^{16}$O$/^{17}$O and $^{16}$O$/^{18}$O ratios compared with these authors 
by $-225\pm 500$ and $-500\pm 620$, respectively. This figure is found for all carbon stars types. We verified
that differences in the stellar parameters adopted for a specific star are not enough to explain the discrepancy
in the measured ratios. Differences in the model atmospheres 
used cannot account for this systematic difference either. In fact, in our analysis the atmosphere models used by 
H87 would imply the derivation of higher oxygen isotopic ratios. H87 adopted the stellar parameters 
derived in \citet{lam86} and used an old grid of unpublished C-rich atmosphere models \citep[see also][]{lam86}.
Thus, we believe that the main reason of the discrepancy is in the difference
between the molecular line lists used, in particular for the CO and CN molecules. In H87 the CO molecular
constant of \citet{dal79} and the log $gf$ values of \citet{cha83} were employed. Unfortunately, we cannot confirm
this because H87 do not indicate the specific C$^{17,18}$O lines used in their analysis. On the other hand, we have four 
stars in common with \citet{dom86}, all of them of SC-type: \object{GP Ori}, \object{FU Mon}, \object{CY Cyg} and \object{RZ Peg}. 
The $^{16}$O$/^{17}$O ratios
derived by these authors are systematically lower than our values, although in \object{FU Mon} and \object{CY Cyg} they agree within the
error bars. Nevertheless, the reason for the discrepancy is completely due to the systematically 
lower T$_{eff}$ values adopted (up to 500 K) by \citet{dom86} compared to ours. Finally, we have few stars in 
common with the recent work by \citet{hin16}. The ratios derived in 
the stars in common marginally agree within the error bars but the comparison in this case is not straightforward because these
authors use the curve-of-growth method in the analysis instead of the spectral synthesis used here.

%
%
\begin{table*}
\caption{Carbon, nitrogen, and oxygen isotopic ratios for AGB carbon stars}             
\label{table:3}      
\centering                          
\begin{tabular}{lcccc}        
\hline\hline                 
Star & $^{12}$C$/^{13}$C\tablefootmark{a} & $^{14}$N$/^{15}$N\tablefootmark{b}& $^{16}$O$/^{17}$O\tablefootmark{c}& 
$^{16}$O$/^{18}$O\tablefootmark{c}\\
\hline 
N-type &    &      &              &       \\
       &    &      &              &       \\
\object{AQ Sgr} & 52 & 1230 & 860$\pm 150$(4)& 763$\pm 280$(8)\\
\object{BL Ori} & 57 & 3700 & 625$\pm 235$(3)& 1094$\pm 600$(7)\\
\object{R  Lep} & 62 & --   & 1730$\pm 700$(6)& -- \\
\object{RT Cap} & 59 & --   & 650$\pm 250$(5)& 735$\pm 220$(6)\\
\object{RV Cyg} & 74 & --   & 610$\pm 190$(6)& 606$\pm 300$(6)\\
\object{S Sct}  & 45 & --   & 784$\pm 176$(3)& 727$\pm 114$(4)\\
\object{ST Cam} & 61 & 1250 & 925$\pm 230$(6)& 1290$\pm 380$(4)\\
\object{TU Gem} & 59 & --   & 1562$\pm 470$(5)& -- \\
\object{TW Oph} & 65 & --   & 830$\pm 400$(5)& 980$\pm 370$(5)\\
\object{TX Psc} & 42 & 1040 &1200$\pm 410$(6)& 660$\pm 370$(7)\\
\object{U Cam}  & 97 & 2000 &1095$\pm 300$(4)& 500$\pm 200$(3)\\
\object{U Hya}  & 32 & --   &1800$\pm 230$(2)& 1220$\pm 100$(5)\\
\object{UU Aur} & 50 & 1100 & 930$\pm 380$(4)& 875$\pm 230$(5)\\
\object{UX Dra} & 26 &  --  & 820$\pm 400$(5)& 560$\pm 40$(5)\\
\object{V460 Cyg}&61 & 4600 & 870$\pm 360$(6)& 562$\pm 300$(5)\\
\object{V Aql}  & 82 & 1800 & 1110$\pm800$(4)& 700$\pm 0$(2)\\
\object{VY UMa} & 44 &  --  & 820$\pm 145$(5)& 725$\pm 450$(7)\\
\object{W CMa}  & 53 &  --  &1060$\pm 250$(4)& 830$\pm 0$(3)\\  
\object{W Ori}  & 79 & 4280 &1200$\pm 500$(6)& 850$\pm 180$(4)\\
\object{Y Hya}  & 82 & --   &850$\pm 270$(6)& 1100$\pm 70$(3)\\
\object{Y Tau}  & 58 & 880  &1250$\pm 300$(6)&1224$\pm 800$(6)\\
\object{X Cnc}  & 52 & 3330 &1045$\pm 165$(7)&1600$\pm 150$(3)\\
\object{Z Psc}  & 55 & 1320 &695$\pm 140$(6)& 732$\pm 280$(3)\\
       &    &      &               &     \\
SC-type&    &      &               &      \\
       &    &      &               &      \\
\object{CY Cyg} & 6  & --   & 620$\pm 50$(3)& 770$\pm 500$(2)\\
\object{FU Mon} & 27 & --   & 535$\pm 100$(7)& 400$\pm 10$(6)\\
\object{GP Ori} & 35 & 660  & 550$\pm 190$(6)& 682$\pm 140$(6)\\
\object{RZ Peg} & 12 & --   & 862$\pm 380$(5)& 890$\pm 400$(5)\\
\object{WZ Cas} & 5  & 640  & 504$\pm 130$(6)& 1250$\pm 350$(7)\\
       &    &      &                &     \\
J-type &    &      &                &     \\
       &    &      &                &     \\
\object{R Scl}  & 19 & --   & 300            & 500 \\
\object{RY Dra} & 3.6& --   & 385$\pm 250$(4)& $>200$\\
\object{T Lyr}  & 3.2& --   & 450$\pm 60$(4) & --\\
\object{VX And} & 13 & 900  & 770            & 650$\pm 170$(3)\\
\object{Y Cvn}  & 3  & 3200 & 270$\pm 55$(4) & $>300$\\
\hline                                   
\end{tabular}
\tablefoot{
\tablefoottext{a}{Carbon isotope ratios from \citet{lam86}, \citet{abi02}, and \citet{hed13}.}
\tablefoottext{b}{Nitrogen isotope ratios from \citet{hed13}.}
\tablefoottext{c}{The number in parenthesis indicates the number of $^{12}$C$^{17,18}$O lines used, respectively.
The dispersion found among the different lines is indicated when more than two lines are used.}
}
\end{table*}

\section{Results and discussion}

 Table 3 shows the oxygen isotopic ratios derived in our sample of AGB carbon stars. The number in parentheses 
indicates the dispersion around the mean value
when more than two $^{12}$C$^{17,18}$O lines were used. The table also shows
the carbon and nitrogen isotope ratios derived for the same stars. In some stars the carbon isotope ratio shown 
is that from \citet{lam86} as our re-analysis here find
only minor differences in this ratio. Note that all the CNO isotope ratios have been derived using a homogeneous set of stellar 
parameters and spectral analysis tools. Therefore, to our knowledge, they constitute the first homogeneous sample of CNO isotopic 
ratios measured in carbon stars. The two oxygen ratios derived cover a similar range:
$300\leq^{16}$O$/^{17}$O$\leq 1800$ and  $200\leq^{16}$O$/^{18}$O$\leq 1600$, being less
wide than those found by H87. For N-type stars the average ratios are $^{16}$O$/^{17}$O$=1057\pm460$ and 
$^{16}$O$/^{18}$O$=870\pm280$. Considering the error bars, these average values are those theoretically 
predicted after the FDU for low-mass stars (see Table 1). Thus, although there are a small number of outlier stars 
deviating from these average values (see below), we may conclude that the oxygen ratios in normal
AGB carbon stars are not significantly altered during
the AGB evolution. This nicely agrees with standard AGB stellar model predictions 
\citep[see e.g.,][and also Table 1]{cri15}. Interestingly, SC- and J-type stars show systematically lower $^{16}$O$/^{17}$O ratios
than normal N-type stars: $614\pm 140$ and $435\pm200$ for SC- and J-type stars, respectively 
(J-type stars also showing lower $^{16}$O$/^{18}$O ratios). This confirms the findings by H87 in their sample of J-type 
stars. SC- and J-type stars have other chemical {\it anomalies} \citep[Li, F, $^{12}$C$/^{13}$C, and $^{14}$N$/^{15}$N, see]
[for details]{wal98,abi03,abi15} to which we have now add the oxygen ratios, making these types of AGB 
carbon stars very peculiar ones.  

%
   \begin{figure}
   \centering
   \includegraphics[width=9cm]{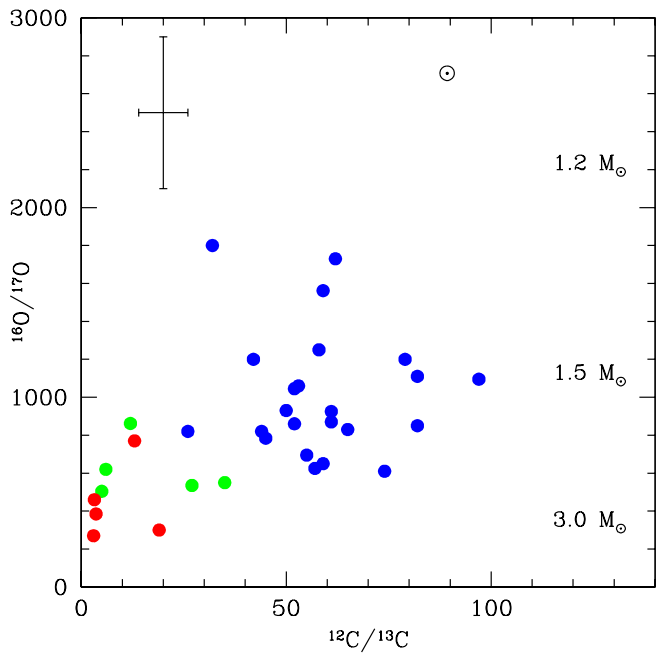} 
      \caption{Derived $^{16}$O$/^{17}$O vs. $^{12}$C$/^{13}$C ratios in the carbon stars
of different spectral types: N-type (blue), SC-type (green) and J-type (red). The position
of the stellar masses on the figure approximately indivcates the value of the predicted $^{16}$O$/^{17}$O ratio
for each mass after the FDU for solar metallicity stellar models. The operation of extra mixing
on the AGB phase would place a star in the region $^{12}$C$/^{13}$C$<30$ and $^{16}$O$/^{17}$O$>1500$ (see Figure 7). 
The solar ratios are marked for a guide.
A typical error bar is shown.}
         
   \end{figure}

%
   \begin{figure}
   \centering
   \includegraphics[width=9cm]{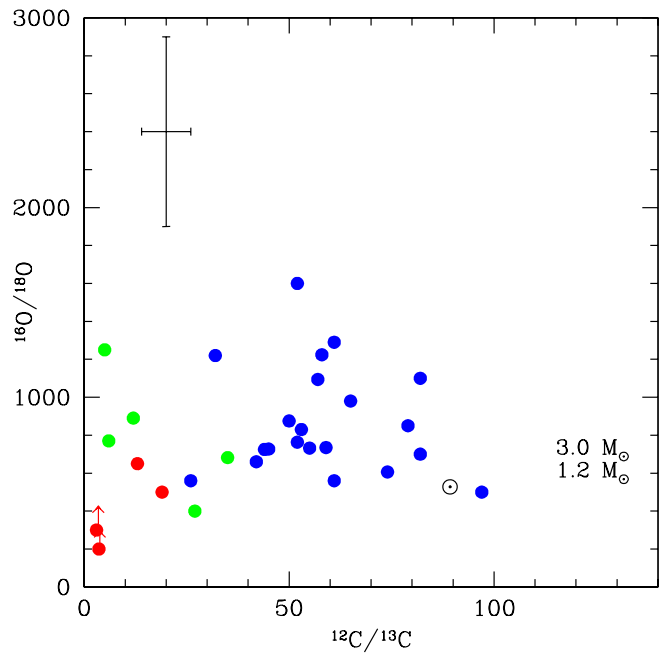} 
      \caption{As figure 4 for the derived $^{16}$O$/^{18}$O vs. $^{12}$C$/^{13}$C ratios. The 
operation of extra mixing on the AGB phase would place a star in the
region $^{12}$C$/^{13}$C$<30$ and $^{16}$O$/^{18}$O$>2000$ (see Figure 7). A typical error bar is shown.}
         
   \end{figure}

Figures 4 to 6 show the $^{16}$O$/^{17}$O$/^{18}$O vs. $^{12}$C$/^{13}$C ratios found in the different types of
carbon stars. Since the $^{12}$C$/^{13}$C ratio increases continuously in the envelope due to the
TDU episodes, this ratio may be considered as a proxy of the evolution along the AGB phase. From these figures,
it is evident that no trend with the carbon ratio exists, suggesting again that  the superficial oxygen ratios remain almost 
unaltered during the AGB evolution. In the figures, the approximate values of the oxygen 
ratios predicted during the AGB phase for several stellar masses at solar metallicity are also shown. Despite the large
error bars, most of the observed ratios clearly lie in the range of theoretical expected values for stars 
within $\sim 1.5-3$ M$_\sun$, in agreement 
with the mass range predicted for the formation of an AGB carbon star \citep[e.g.,][]{str03,kar100,cri15}. This can be
better seen  in Figure 6 since theoretical predictions for the $^{17}$O$/^{18}$O ratio are much more sensitive to the stellar mass.
Furthermore, this ratio has a lower uncertainty because errors in the stellar parameters affect 
both the $^{16}$O$/^{17}$O and $^{16}$O$/^{18}$O ratios in the same way and cancel out when deriving the $^{17}$O$/^{18}$O ratio.

Figure 4 shows, in particular, that stars of SC and J types have, on average, lower $^{16}$O$/^{17}$O ratios than ordinary 
carbon stars, and also have very low ($<30$) $^{12}$C$/^{13}$C. As mentioned in Section 1, carbon ratios $\leq 30$ in carbon stars
are difficult to explain by AGB standard models. On the other hand, two N-type stars (\object{TU Gem} and \object{Y Tau}) have high $^{16}$O$/^{17}$O but
normal $^{12}$C$/^{13}$C. In Figure 5, some N-type stars also show  a large $^{16}$O$/^{18}$O$>1000$ ratio.
These large $^{16}$O$/^{17}$O ratios cannot be explained by standard AGB models either and might require a
non-standard mixing process. Such large ratios in the AGB phase would be expected, nevertheless,
in stars with initial masses between $\sim 1.1$ and 1.3 M$_\sun$ at solar metallicity (see Table 1). For such low
stellar masses however, the envelope is so small that TDUs are very inefficient and the stars would never become a carbon
star, which is against the evidence derived for our stars. Note however, that the derived $^{17}$O$/^{18}$O ratio (with a lower
uncertainty) in these two N-type stars, is compatible with theoretical predictions for stars with 
M$\geq 1.5$ M$_\sun$ (see Table 1 and Figure 6).

%
   \begin{figure}
   \centering
   \includegraphics[width=9cm]{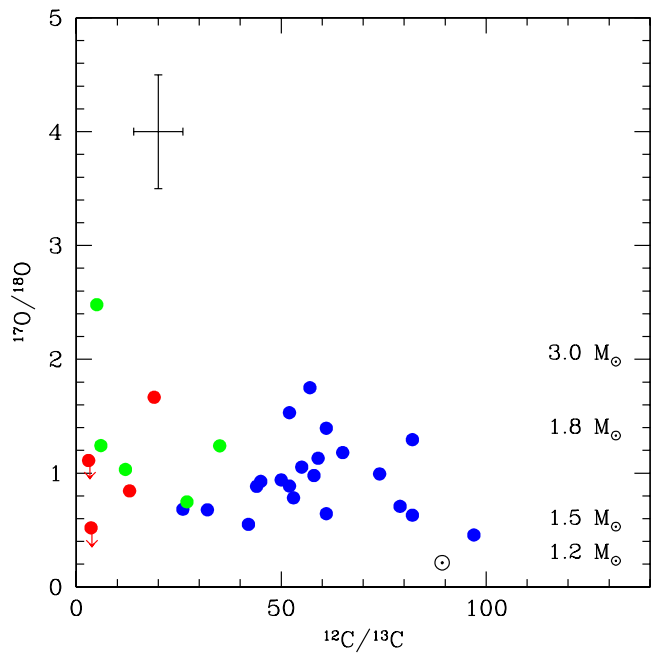} 
      \caption{As figure 4 for the derived $^{17}$O$/^{18}$O vs. $^{12}$C$/^{13}$C ratios. Note the
extreme sensitivity of the $^{17}$O$/^{18}$O to the stellar mass after the FDU. Most of the measured ratios
lie within the values expected for masses 1.5-3 M$_\odot$. The existence of extra mixing on the AGB phase 
would place any star in the region $^{12}$C$/^{13}$C$<30$ and $^{17}$O$/^{18}$O$>4$. A typical error bar
is shown.}
         
   \end{figure}

\subsection{Comparison with theoretical low-mass AGB stellar models}

As mentioned in Section 1, it is generally accepted that the FDU alone cannot account for the CNO isotope ratios  measured in
RGB stars with M$< 2$ M$_\sun$ and that an extra mixing process must  be at work in the upper RGB. In particular, the $^{12}$C$/^{13}$C ratio 
in these stars is substantially lower than predicted by stellar models after the occurrence of the FDU. Values of $^{12}$C$/^{13}$C$< 20$ are 
commonly found in low-mass RGB stars brighter than the so-called RGB bump. The observed carbon ratios are even lower than 
approximately ten in 
metal-poor RGB stars. In principle, also nitrogen and oxygen ratios may be altered by extra mixing in RGB stars, although there is
little observational evidence of this \citep[see e.g.,][]{abi12}. On the other hand, the existence of 
extra mixing in AGB stars is still under discussion 
(see Section 1).  The actual physical process capable to drive  
efficient mixing below the convective envelope in giant stars is still largely unknown. In any case, to explore this possibility we have 
constructed AGB evolutionary models including extra mixing using the latest release of the FUNS code
\citep{stra06,pie13} for two representative stellar masses (1.5 and 2 M$_\sun$ with Z$=0.014$, Y$=0.27$). Except for 
the $^{17}$O$(p,\alpha)^{14}$N \citep{bru16}, all the rates of the other reactions involved in the H burning are from 
\citet{ade11}. The various uncertainties affecting stellar models and the related nucleosynthesis results, including those due
to the treatment of mixing, are discussed in \citet{str14}. We have assumed solar initial CNO ratios. Despite the fact that our
stars are considerably younger than the Sun, current observations in the local ISM show average CNO ratios compatible with the
solar ones, albeit with a significant dispersion \citep{mil05,ada12,rit15,nit12}.

Then the idea is to study the effects of the extra mixing models on the CNO
isotope ratios and compare the results with the observed ones exploiting the fact that we have measured the {\it three} 
isotope ratios in some of our stars. We have modeled this following the procedure in \citet{nol03}. Starting from the RGB-bump, 
we have switched on an artificial extra mixing process extending from the convective boundary down to an assumed maximum 
temperature T$_{max}$ \citep[see][for details]{dom04}.
The mixing velocity has been set to $100$ cm/s, which is of the same order of magnitude as the velocity estimated for
thermohaline mixing \citep[see e.g.,][]{lat15} and magnetic buoyancy \citep{nuc14}, while it is an order of magnitude larger than that 
expected for rotational induced mixing \citep{zah13}. We checked that the CNO isotopic ratios at the stellar surface do not 
depend on the assumed 
mixing velocity within a reasonable range, but mainly on T$_{max}$. Extra mixing processes have been activated from the 
RGB bump to the RGB tip
(T$_{max} = 22$ MK). In this parametrized 
extra mixing model, we have assumed that the T$_{max}=22$ MK remains constant up to the RGB tip. This maximum temperature 
approximately corresponds to the temperature of the envelope/core interface at the epoch of the RGB bump. Later on, however, 
the H-burning shell becomes hotter and one may suppose that T$_{max}$ could be larger. In that case a 
lower $^{12}$C$/^{13}$C would be attained at the RGB tip. Figure 7 shows the evolution of the
main CNO abundances and isotope ratios along the RGB and AGB phases for the 1.5 (black line) and 2 M$_\sun$ (red line) models.
For clarity, the evolutionary timescale (horizontal axis) has been re-scaled to the final time in the computation $t_f$, namely, 
$2.91$ and $1.196$ Gyr  for the 1.5 and 2 M$_\sun$ models, respectively.  The FDU occurs at $\log t/(t-t_f) \sim 0.8$ (1.5 M$_\sun$) 
and 0.6 (2 M$_\sun$), while the RGB bump takes place at $\log t/(t-t_f) \sim 1.2$ and 0.9, respectively. 
The $^{12}$C depletion and the corresponding $^{14}$N enhancement are clearly shown as well as the decreases of
the $^{12}$C$/^{13}$C, C/O, and C/N ratios. The effect of the $^{17}$O enhancement on the $^{16}$O$/^{17}$O ratio is also evident.
Due to the large vertical scale, the increase of $^{14}$N$/^{15}$N and $^{16}$O$/^{18}$O ratios, as caused by the FDU, is less
evident.

  \begin{figure*}
\centering
            \includegraphics[width=16cm]{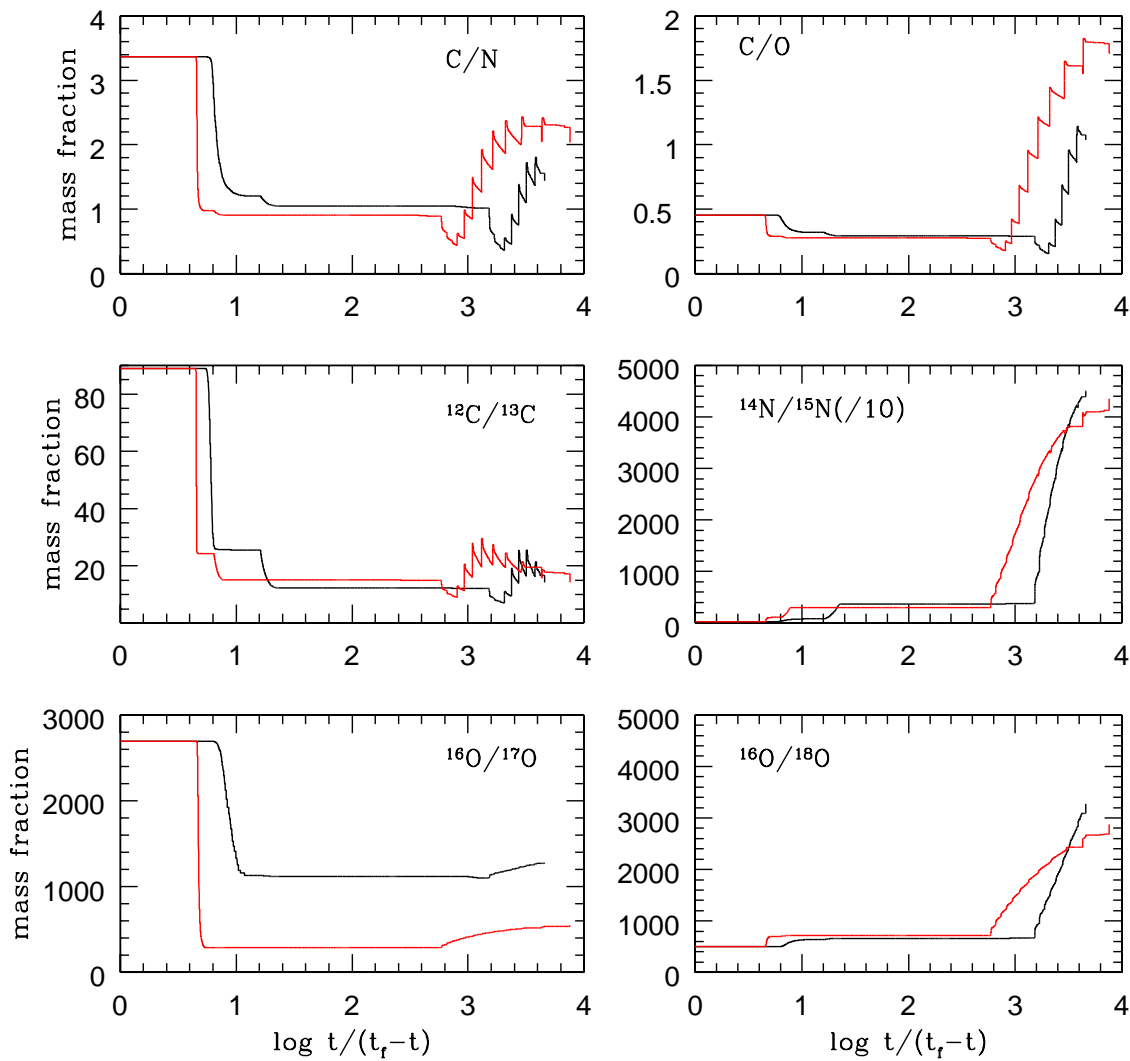}
      \caption{Evolution up to the AGB tip of C, N, and O elemental and isotopic ratios in models with 1.5 M$_\sun$ (black line) 
and  2 M$_\sun$ (red line). The first dredge up occurs at $\log t/(t-t_f) \sim 0.9$ (1.5 M$_\sun$) and 0.6 (2 M$_\sun$), 
while the RGB bump takes place at $\log t/(t-t_f) \sim 1.2$ and 0.8, respectively. The
the combined actions of the third dredge up and the AGB extra mixing arise at  $\log t/(t-t_f) > 3.2$ and  2.8. Note 
the reduced scale in the $^{14}$N$/^{15}$N ratio.}

   \end{figure*}  
In addition to the RGB extra mixing, an AGB extra mixing extending down to T$_{max} = 40$ MK has been activated 
since the first thermal pulse to the AGB tip. This can be seen in Figure 7 near log $t/(t_{f}-t)\sim 3$. Due to extra mixing the star now 
needs more TDU events to become a carbon star (C/O$>1$) than in the standard case. This has a dramatic effect on the
$^{12}$C$/^{13}$C ratio in the envelope which is kept below $\sim 30$ throughout the AGB phase (see Figure 7). This result is
well known: to explain the low $^{12}$C$/^{13}$C ratios measured in a non-negligible fraction of Galactic AGB carbon stars,
extra mixing both in the RGB and AGB phases is needed (see references in Section 1). Concerning the oxygen ratios, 
the $^{16}$O$/^{17}$O would increase 
since at T$_{max}=40$ MK the CNO cycle equilibrium value of this ratio is approximately 1870 and, 
as a consequence of the AGB extra mixing, the material with such a high $^{16}$O$/^{17}$O ratio in the most internal portion of 
the H-burning shell is mixed with the envelope material that is characterized by a low $^{16}$O$/^{17}$O ratio after the FDU. 
This effect is weaker for the 1.5 M$_\sun$ model, because the $^{16}$O$/^{17}$O ratio after the FDU is higher (see Figure 7). 
AGB extra mixing would, nevertheless, increase the $^{16}$O$/^{17}$O ratio by less than a factor of two with respect to
the value remaining after the FDU. 

On the contrary, an AGB extra mixing would have a huge impact on the $^{16}$O$/^{18}$O ratio due to the 
substantial $^{18}$O depletion occurring in the CNO burning region: when C/O $>1$ this ratio is higher than 
2000 for the two stellar
models shown in Figure 7. Furthermore, AGB carbon stars affected by extra mixing would reach extreme 
$^{14}$N$/^{15}$N ratio values ($>10^4$, see Figure 7 and Table 4). In summary, carbon stars with a $^{12}$C$/^{13}$C ratio lower
than 30 can be explained by an 
AGB extra mixing, but this would simultaneously typically imply $^{16}$O$/^{17}$O$>1000$,  $^{16}$O$/^{18}$O $>2000$,  
and $^{14}$N$/^{15}$N$>10^4$. The stars in Figures 4 to 6 with such a
low  $^{12}$C$/^{13}$C ratio do not show these extreme oxygen and nitrogen ratios and, vice versa, the few carbon 
stars in our sample with such large $^{16}$O$/^{17}$O$/^{18}$O and/or $^{14}$N$/^{15}$N ratios 
(see Figures 4 to 6 and Table 3), have normal $^{12}$C$/^{13}$C ratios ($>40$). For these stars, there is no
stellar scenario on the basis of the hydrostatic H-burning capable of explaining the measured CNO ratios.
This constitutes a challenge for stellar physics. Nevertheless, a way to escape  this puzzle is 
the possibility (although improbable) mentioned above 
that the stars with high O ratios have a very low mass ($<1.3$ M$_\sun$). Another alternative scenario would 
involve the assumption that these stars 
were born with extremely large oxygen ratios. According to our simulations with the FUNS code initial values $^{16}$O$/^{17}$O$>4000-5000$ and/or 
$^{16}$O$/^{18}$O$>1000$ would be required. Despite there being
a significant dispersion in these ratios at a given galactocentric radius, such large ratios are not observed in the local
ISM (within $\sim 1$ kpc around the Sun) \citep{wil94,kah96,nit12}. Table 4 summarizes the values of the CNO isotopic ratios reached 
at the main mixing events when extra mixing is included in both the RGB and AGB phases for the stellar
models shown in Figure 7. 

\begin{table*}
\label{fdu} 
\centering 
\caption{Changes of the CNO isotopic ratios after the major deep mixing events including extra mixing.} 
\begin{tabular}{c c c c c c} 
\hline\hline 
       & $^{12}$C/$^{13}$C  & $^{14}$N/$^{15}$N & $^{16}$O/$^{17}$O  & $^{16}$O/$^{18}$O & $^{17}$O/$^{18}$O \\
\hline
\hline 
     M=1.5 M$_\odot$ &  & & & & \\
\hline
FDU     & 26 & 781 & 1095 & 637 & 0.58 \\
RGB tip & 12 & 3666 & 1120 & 662 & 0.59 \\
AGB tip & 14 & 45124 & 1277 & 3271 & 2.56 \\
\hline
M=2.0 M$_\odot$ &  & & & &   \\
\hline
FDU     & 24 & 1072 & 295 & 701	& 2.38 \\
RGB tip & 15 & 2965 & 296 & 718 & 2.43 \\
AGB tip & 14 & 43550 & 551 & 2876 & 5.22 \\
\hline
\end{tabular}
\end{table*}

On the other hand, we have not found any correlation between the $^{16}$O$/^{17}$O$/^{18}$O ratios and the neutron exposure later measured
by the abundance [hs/ls] ratio (see Section 1). To derive the [hs/ls] ratio we have used the s-element abundances derived in many of 
the stars studied here according to \citet{abi02}. This is at odds with suggestions in H87, but indeed, no such correlation is 
predicted by current AGB stellar models.

\subsection{Carbon stars and pre-solar oxide grains} 

Finally, we compare the derived oxygen ratios with those in pre-solar oxide
grains. In the cosmochemistry community the isotopic ratios are usually the inverse of
the astronomical convention. In this section, we adopt their convention to allow
comparisons to the literature data. As mentioned in Section 1, the majority of pre-solar oxides are
 believed to have formed in O-rich RGB and AGB stars. We point out that due to the intrinsic C-rich 
nature of our objects, a direct comparison of the oxygen ratios derived in AGB carbon stars with those
in oxide grains should be avoided. However, in the standard evolutionary framework, carbon stars are
the natural descendants of the O-rich AGBs (of spectral types M, MS and S). Therefore, such comparison might
give some indication on whether or not the surface oxygen ratios have undergone some evolution from the O- to the C-rich phase
along the AGB evolution. In fact, there is observational evidence for the presence of spectral features produced
by oxide and silicate grains in the circumstellar envelopes of some intrinsic (i.e., no binaries) carbon stars 
\citep[see e.g.,][]{lit86,kwo93,che11,guz15}. These spectral features seem to have their origin in the outer O-rich shells
of the circumstellar envelope formed when the star was an O-rich AGB. 

Because the typical lifetime of stellar grains in the ISM is estimated to be $\sim 0.5$ Gyr \citep{jon96}, the parent
stars of pre-solar grains recovered now must have ended their lives relatively shortly before 
the solar system formed 4.6 Gyr ago. In contrast, the AGB stars studied here are relatively young ($\sim 1-2$ Gyr), born
much later that the solar system. This means that our stars are, on average, more metal-rich than the AGB stars from which
the pre-solar grains formed. As a consequence, from chemical evolution considerations \citep{pra96}, our stars should have been formed with
different initial oxygen ratios than those in the progenitors of the pre-solar grains, that is, with higher than
solar $^{17}$O$/^{16}$O and $^{18}$O$/^{16}$O ratios. Note that according to standard stellar models, oxygen ratios in the AGB
phase depend mainly on values attained after the FDU, which in turn, depend on the initial oxygen ratios and stellar mass.  
Keeping this in mind, in Figure 8 we compare the oxygen ratios derived for our stars with those in the oxide
grains of types I and II. Also, the oxygen ratios observed in O-rich AGB stars of spectral types M, MS, and S
\citep[open circles]{smi90} and in O-rich Miras \citep[black dotted circles]{hin16} are shown. Excluding the
upper limits set to the $^{18}$O abundance in two J-type stars (red circles), all the oxygen ratios
in carbon stars lie in the range observed in the oxide grains of type I, which are believed to
form in stars with $1.2\leq$M/M$_{\sun}\leq 2$. Note that, the range of oxygen ratios measured here
are similar to those found by \citet{smi90} in O-rich AGBs. This could be interpreted
as an indication of little evolution of the stellar surface ratios from the O- to the C-rich phase, in agreement with
theoretical predictions.  It is also apparent that they distributed in a small range, showing little spread. 
This is compatible with the narrow range 
in mass at which the formation of a carbon star is expected at solar metallicity 
($1.5-3$ M$_\odot$). In contrast, the large spread in the $^{17}$O$/^{16}$O 
ratios observed in O-rich AGB stars \citep[mainly in the Miras from][]{hin16} agrees with 
the idea that these stars descend from a much wider range of initial masses (see Table 1). On the other hand, no carbon 
star in our sample is found with significant $^{18}$O enhancement (larger than the solar $^{18}$O$/^{16}$O ratio), in contrast with that
observed for the Mira stars in \citet{hin16} (black dotted circles in Figure 8). These authors interpret 
this figure as a clear indication of the chemical evolution in the Galaxy since the $^{18}$O$/^{16}$O ratio should increase 
with time. However, our stars should 
have similar ages to those in \citet{hin16}, but none of them show such $^{18}$O enhancement. 
This finding is difficult to understand. On the other hand, from Figure 8 it is evident that
only a few O-rich AGB stars show an $^{18}$O depletion as large as
that found in oxide grains of type II ($^{18}$O$/^{16}$O$<<10^{-3}$). These grains are believed to form in low-mass RGB and AGB stars
in which extra mixing has taken place \citep{was95,pal13}. As discussed before, in general
the oxygen ratios found in our stars combined with the observed carbon and nitrogen
ratios, leave little space for the occurrence of extra mixing, at least in ordinary N-type AGB carbon stars. Since, on average,
the stars in Figure 8 should have larger metallicity than the stars origin of the
pres-olar grains, \citet{hin16} interpreted this figure as evidence that extra mixing is a rare 
phenomenon in metal-rich stars. The oxygen ratios derived here for AGB carbon stars support this conclusion.
Obviously, more observational and theoretical studies are
needed to elucidate this.  

%
   \begin{figure}
   \centering
   \includegraphics[width=9cm]{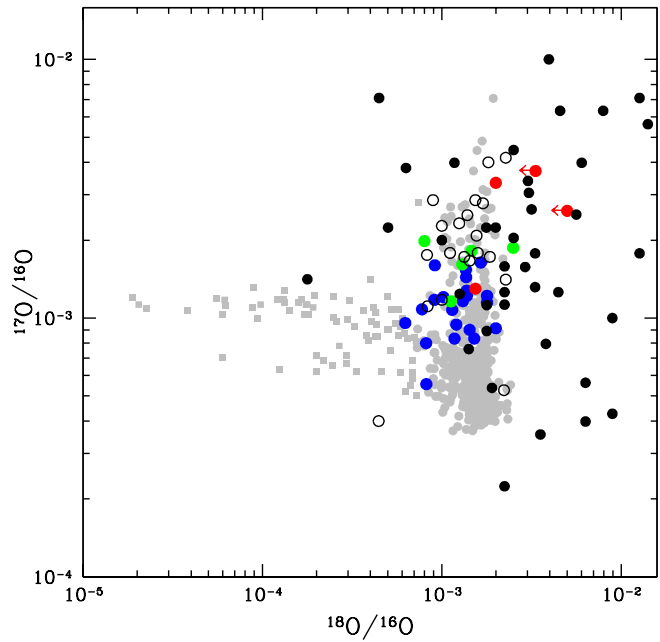} 
      \caption{Comparison of $^{17}$O$/^{16}$O and $^{18}$O$/^{16}$O ratios in AGB stars with
those derived in oxide pre-solar grains of type I (gray dots) and II (gray squares) as defined by
\citet{nit94}, taken from the database http://presolar.wustl.edu. As figure 1, color symbols are the AGB carbon stars 
analyzed here. Open black circles: O-rich AGB stars of types M, MS and S from \citet{smi90}; solid black 
circles: O-rich Miras from \citet{hin16}. Error bars in the pre-solar grains are smaller than
the symbol size. Note the logarithmic scale.}
        
   \end{figure}

\section{Conclusions}

 We have presented the first homogeneous analysis of carbon, nitrogen and
oxygen isotope ratios in a sample of Galactic AGB carbon stars. For most of the
stars, the nitrogen and oxygen ratios found are compatible with the predicted values after the
occurrence of the first dredge-up suggesting that they are not significantly altered
along the AGB evolution. This agrees with the predictions of standard theoretical low-mass (1.5-3 M$_\odot$) 
AGB models. Furthermore, the $^{12}$C$/^{13}$C ratios are typically larger than 40 as expected from the operation of
the third dredge-up episodes. Some stars (mostly of J- and SC-types) showing  
$^{12}$C$/^{13}$C$<30$, would require extra mixing in both the RGB and AGB phases. However, these stars
do not show the corresponding high $^{16}$O$/^{17}$O ($>1000$), $^{16}$O$/^{18}$O ($>2000$) 
and extreme $^{14}$N$/^{15}$N ($>10^4$) ratios expected when extra mixing takes place. Conversely, a
few carbon stars with high $^{16}$O$/^{17}$O$/^{18}$O do not show low $^{12}$C$/^{13}$C ratios. For these
stars, the observed CNO ratios cannot be explained on the basis of the hydrostatic H-burning.
Globally, the derived CNO isotopic ratios  would suggest that
the operation of extra mixing along the AGB phase is quite a rare phenomenon, at least for 
solar metallicity (N-type) carbon stars.

On the other hand, the $^{16}$O$/^{17}$O$/^{18}$O ratios derived here are very similar to those found in
pre-solar, type I oxide grains, supporting the idea that these grains were formed in AGB stars
during their O-rich phase. Considering O- and C-rich AGB stars together, we find almost no stars with a $^{18}$O$/^{16}$O ratio as
low as those observed in oxide grains of type II. This casts some doubt on the possible formation of these
pre-solar grains in AGB stars.

\begin{acknowledgements}
      Part of this work was supported by the Spanish MEC grants
      AYA-2011-22460 and AYA2015-63588-P within the European Founds for
Regional Development (FEDER). NSO/Kitt Peak FTS data used here 
      were produced by NSF/NOAO. Some of the data shown here are part
of R.P. Hedrosa's Ph.D thesis. We thank S. Cristallo and P. de Laverny
for revising this Ph.D. work. ID thanks the support of the National Science Foundation 
under Grant No. PHY-1430152 (JINA Center for the Evolution of the Elements). 
\end{acknowledgements}

%
   \bibliographystyle{aa} 
   \bibliography{ox2.bib} 
%
\end{document}